\documentstyle[12pt,aasms4]{article}

\begin{document}

\title{FUSE 900--1200\AA\ Spectroscopy of AM Her\footnote{Based on
observations made with the NASA-CNES-CSA Far Ultraviolet Spectroscopic
Explorer.  FUSE is operated for NASA by the Johns Hopkins University 
under
NASA contract NAS5-3298} }

\author{J.B. Hutchings}
\affil{Herzberg Insitute of Astrophysics,
National Research Council of Canada,\\ Victoria, B.C. V8X 4M6, Canada\\
john.hutchings@nrc.ca}

\author{A.W. Fullerton}
\affil{Dept. of Physics and Astronomy, University of Victoria\\
P.O. Box 3055, Victoria, B.C. V8W 3P6, Canada}
\affil{Dept. of Physics and Astronomy, Johns Hopkins University\\
3400 N. Charles St., Baltimore, MD 21286\\
awf@pha.jhu.edu}

\author{A.P.\ Cowley\altaffilmark{1}, P.C.\ Schmidtke\altaffilmark{1}}
\affil{Dept. of Physics \& Astronomy, Arizona State University,
Tempe, AZ, 85287-1504\\~anne.cowley@asu.edu; ~paul.schmidtke@asu.edu}

\begin{abstract}

Spectra of the magnetic white dwarf binary AM Her were obtained with the
Far Ultraviolet Spectroscopic Explorer (FUSE) during three consecutive
spacecraft orbits.  These observations were split into 19 spectra of about
5 min duration (0.025P binary phase) partially covering the binary orbit.
We report the phase-related changes in the far ultraviolet continuum light
curve and the emission lines, noting particularly the behavior of O VI.
We discuss the fluxes and velocities of the narrow and broad O VI emissions. 
We find the FUV light curve has maximum amplitude at $\sim$1000\AA,
although at shorter wavelengths the continuum may be strongly affected by
overlapping Lyman lines.  Weak, narrow Lyman absorption lines are present.
Their velocities don't appear to vary over the observed orbital phases, 
and their mean value is consistent with the systemic velocity. 

\end{abstract}


\section{Introduction}

AM Her, originally classified as an irregular variable and associated with
the X-ray source 3U 1809+50, was discovered in 1976 to be a short period
binary related to the cataclysmic variables (CVs) (Szkody \& Brownlee
1997, Cowley \& Crampton 1977, Priedhorsky 1977).  Through its 3.1-hr
orbital period, AM Her's brightness varies continuously, with
short-term flickering superimposed on an orbital light curve with a range
of $\Delta$$V\sim0.7$ mag.  Independently, Tapia (1977) found the same
orbital period from polarization variations indicating that the white
dwarf in the system has a very strong magnetic field ($B\sim 2\times10^8$
gauss) and rotates synchronously with the orbit.  The magnetic field
constrains the mass lost from the secondary star to flow along a stream
directly onto the white dwarf rather than into an accretion disk as in
the non-magnetic CVs.

The optical spectrum shows strong emission lines of H, He~I, and He~II,
along with weaker lines of other ions such as N~III, C~III, C~II, and Ca~II,
all showing large velocity variations.  The strong lines show both broad
and narrow components which clearly arise in different regions since their
phasing and velocity amplitudes differ considerably (e.g. Crampton \&
Cowley 1977, Crosa et al. 1981, and many more recent studies).  TiO 
bands from the M4~V secondary star are detected when the system is in its 
faint  (``low") state (e.g. Young, Schneider, \& Shectman 1981).

Even early observations showed light curves and times of minima in $V$,
$U$, and X-ray bands to be quite different (e.g. see Fig.\ 6 of Crampton
\& Cowley, 1977).  Subsequent observations in other wavelengths have revealed
additional complex patterns of orbital behavior, much of which can be
interpreted as a result of the strong magnetic field.  G\"ansicke et al.\
(2001) attribute the $V$-band light curve to cyclotron emission arising
near the accreting magnetic pole, the small $B$-band variation to the
accretion stream, and the hard X-ray variation to the changing view of the
hot plasma at the shock where stream impacts the white dwarf.  Because
times of minima depend on the wavelength of observation, there is some
uncertainty about what ephemeris to use.  In this paper we adopt the
well-determined period given by Heise \& Verbunt (1988) and a phase based
on our time of FUV minimum light.  In other published work often the
adopted T$_0$ is the time of minimum $V$ light which corresponds
approximately to the superior conjunction of the white dwarf (e.g.
G\"ansicke et al. 1998, Southwell et al. 1995).

Although AM Her usually remains in its `high' state ($V\sim$13 mag),
from time to time it drops into a `low' state ($V$ below 15 mag) which may
last weeks or months (e.g. Mattei 1980).  A plot of the long-term light
curve from 1977 to 1998 is shown in Fig.\ 1 of Hessman, G\"ansicke, \&
Mattei (2000), which nicely illustrates the high and low states of AM Her.
Although the system underwent a rather prolonged low state a few years ago
(e.g. de Martino et al. 1998), the FUSE data described in the present
paper were obtained during a normal `high' state, with magnitude close to
$V\sim13.3$, as is shown by the on-line AAVSO database.

AM Her has become the proptotype of the class of magnetic CVs, commonly
called ``polars", in which the rotation period of the white dwarf is
locked to the orbital period.  An excellent summary of the properties of
these systems is given by Warner (1995).

\section{FUSE Observations}

Observations of AM Her were made with the Far Ultraviolet Spectroscopic
Explorer (FUSE) on 2000, June 12-13 using the large science aperture and
recording the data in time-tag mode.  The observations covered three FUSE
orbits, and the archival processing (Calfuse 1.6.9) produced three spectra
- one for each FUSE orbital window.  We also re-extracted the spectra with
Calfuse 1.9.9, with the time-tagged data split into
19 individual spectra of duration
0.0033 days each (4.75 min, or 0.025P in binary orbital phase).  Because
of earth occultations, there are gaps in the phase coverage so that only
about half the orbit is well sampled.  This includes phases
$\Phi_{UVmin}$=0.28-0.52 and 0.84-0.97, using an ephemeris based on the
period given by Heise \& Verbunt (1988) and T$_0$ corresponding to the
time of minimum far-ultraviolet light (as given in \S 4).  However, the
observed phases do cover two main portions of the orbit, about
180$^{\circ}$ apart, with some phase overlap.  The spectra have sufficient
signal and time resolution to obtain phase-related information.

\section{FUSE Spectrum of AM Her}

Figure 1 shows two spectra obtained from the entire second and third FUSE
orbits.  These spectra cover the phases near minimum and maximum FUV
light.  Stellar emission features which are present include O VI 1032,
1038\AA, He II 1085 \AA, N III 992\AA, C III 977, 1175\AA, and S IV
1073\AA.  These lines are labelled in Fig.\ 1, and their rest wavelengths
are marked.  Note the continuum differences and the changes in the
emission line structure between the two spectra.

Measurements of these lines are given in Table 1.  The HJDs in the table
have been converted from the MJDs provided in the FUSE data header, and
the heliocentric time correction has been added.  Heliocentric velocity
corrections are included in the standard FUSE reductions.  Most of these
lines were previously observed with $ORFEUS~II$ and Hopkins Ultraviolet
Telescope (HUT), but with lower spectral resolution (Mauche \& Raymond
1998, Greeley et al. 1999).

The strong narrow emission lines in the summed spectra are mainly airglow,
except for the narrow component in the O VI lines.  O VI shows both a
broad and a narrow emission component, similar to the structure seen for
the strongest lines in the optical and ultraviolet regions.  There is also
broad emission from the stronger Lyman lines. G\"ansicke et al.\ (1998)
point out that in the UV only the highest excitation lines (N V and Si IV)
show this broad plus narrow line structure, while the lower ionization
features of C II, C III, Si II, and Si III show only the broad component.
In the FUV we similarly find that the C III lines (and probably the Si IV
lines, which are weaker) have only the broad
emission line, although the profiles change significantly with binary phase.

Narrow Lyman absorption lines are seen below 950\AA, especially noticeable
in the spectrum at maximum FUV light.  These were not observed in the
$ORFEUS~II$ spectra, probably because of their lower resolution.  There is
only weak evidence for any H$_2$ absorption, in strong contrast to the
supersoft X-ray binary QR And in which Hutchings et al.\ (2001) found
probable circumbinary as well as interstellar H$_2$ absorption.

Airglow emissions lie in the middle of the N III 992\AA\ and He II
1085\AA\ emission lines, making it difficult to make clean measurements of
these features.  While there are some airglow lines near the O VI lines,
we can make use of the doublet to isolate these fairly well.  The strong C
III 1175\AA\ line is free of airglow and shows very different profile
changes from O VI, as described in \S 5 below.

\begin{deluxetable}{llcccc}
\tablecaption{Spectroscopic Data and Measurements}
\tablehead{\colhead{HJD} &\colhead{FUV\tablenotemark{1}}
&\colhead{Flux } &\colhead{O VI }
&\colhead{O VI } &\colhead{C III }\\
\colhead{2451700+} &\colhead{Phase} &\colhead{980--1090\AA}
&\colhead{peak} &\colhead{broad} &\colhead{broad}\\
& & \colhead{(erg/cm$^2$/sec/\AA)}
&\colhead{(km s$^{-1}$)} &\colhead{(km s$^{-1}$)} &\colhead{(km 
s$^{-1}$)} }
\startdata

8.4956 &0.281 &2.25$\times10^{-13}$  &$-$27, $-$36 &214, --- & 300\nl
8.4989 &0.306 &2.62$\times10^{-13}$  &$-$33, $-$18 & 69, 63  & 430\nl
8.5023 &0.333 &2.94$\times10^{-13}$  &$-$38, $-$47 &159, $-$27 & 425\nl
8.5056 &0.358 &3.09$\times10^{-13}$  &$-$41, $-$53 &252, 97  & 305\nl
8.5089 &0.384 &3.21$\times10^{-13}$  &$-$44, $-$47 &66, 251  & 610\nl
8.5122 &0.410 &3.03$\times10^{-13}$  &$-$47, $-$44 &267, 245 & 355\nl
8.5155 &0.435 &3.18$\times10^{-13}$  &$-$53, $-$44 &110, 126 & 530\nl
8.5676 &0.839 &1.55$\times10^{-13}$  &66, 57 &$-$247, $-$189 & 
$-$180\nl
8.5709 &0.865 &1.45$\times10^{-13}$  &95, 71 &$-$294, $-$348 & 
$-$230\nl
8.5742 &0.890 &1.31$\times10^{-13}$  &69, 66 &$-$172, $-$333 & 
$-$225\nl
8.5775 &0.916 &1.06$\times10^{-13}$  &66, 83 &$-$201, $-$119 & 
$-$280\nl
8.5808 &0.942 &1.03$\times10^{-13}$  &89, 83 &$-$178, $-$134 & 
$-$260\nl
8.5841 &0.967 &0.94$\times10^{-13}$  &81, 80 &$-$152, $-$206 & 
$-$255\nl
8.6385 &0.389 &3.04$\times10^{-13}$  &$-$44, $-$50 &113, 216 & 150\nl
8.6418 &0.415 &3.00$\times10^{-13}$  &$-$53, $-$41 &113, 57  & 250\nl
8.6451 &0.440 &3.20$\times10^{-13}$  &$-$41, $-$50 &101, 112 & 100\nl
8.6484 &0.466 &3.50$\times10^{-13}$  &$-$35, $-$38 &127, --- & 200\nl
8.6517 &0.492 &3.27$\times10^{-13}$  &$-$33, $-$10 &211, 222 & 250\nl
8.6550 &0.517 &3.38$\times10^{-13}$  &$-$33, $-$33 &142, 199 & 300\nl
\enddata
\tablenotetext{1}{Ephemeris: T$_0$($FUVmin$) = HJD 2451708.4594 +
0.128927041E; see text }
\end{deluxetable}

\section{FUV Continuum Changes and Adopted Phases}

There are continuum changes across the whole FUSE range.  We have defined
a FUV phase, using the period of Heise \& Verbunt (1988) and the time of
minimum FUV light from our FUSE data.  The continuum light curve used for
our ephemeris is based on the total signal from the LiF1a channel, after
removal of the strong airglow emissions.  This channel has the best
guiding and signal levels in the FUSE data.  As discussed below, there are
differences within the FUSE band, so we also measured the continuum over
several ranges of wavelengths (see Figure 2, lower panel). 
Although all the continuum flux datasets
are well fit by a sine curve, we have used the fit for 980-1090\AA\ to
define the time of minimum FUV flux and hence our zero phase
($\Phi_{FUVmin}$=0.0).  Table 2 shows the parameters of the fits for
various wavelength ranges.  We adopt:

\centerline{$\Phi_{FUVmin}=0$ = HJD 2451708.4594(11) + 
0.128927041(5)E}

Comparing this to other phase conventions used in the literature, we find
$\Phi_{mag}$ = $\Phi_{FUV}$ + 0.11 (based on the ephemeris given by Heise
\& Verbunt 1988).  Similarly orbital phases, where zero phase is the
superior conjunction of the white dwarf, are $\Phi_{orb}$ = $\Phi_{mag}$ +
0.37 = $\Phi_{FUV}$ + 0.48 (using information from Southwell et al. 1995
and G\"ansicke et al. 1998).  The problem with these alternative zero points
is that they are defined by older data, so here we prefer to use only
our FUV data to avoid possible error accumulation over many years.

    The phasing for our FUV light curve is very similar to that
observed in other ultraviolet and far-ultraviolet studies.  Using IUE
data, G\"ansicke, Beuermann, \& de Martino (1995) found the flux at
$\lambda$1460 brightest at $\Phi_{mag}\sim0.6$ and faintest at
$\Phi_{mag}\sim$0.1 (see their Fig.\ 5).  From HST data, G\"ansicke et
al.\ (1998) observed similar phasing for three different wavelength
regions between 1150\AA\ and 1427\AA, with the amplitude increasing
towards shorter wavelengths (see their Fig.\ 3).  Greeley et al.\ (1999),
using HUT data, found the peak brightness at the Lyman limit occurred at
$\Phi_{mag}\sim0.6$, and Mauche \& Raymond (1998) found the flux at
1010\AA\ also peaked at magnetic phase $\sim$0.6.  Formally, our peak
occurs at $\Phi_{mag}=0.61$.  Thus, all of these independent data sets are
consistent in the phasing of the UV and FUV light curves.

However, in our FUSE data the continuum shows a deeper minimum over the
wavelength range 950-1050\AA\ than outside that range, with the deepest
minimum occuring at $\sim$1000\AA.  There the total flux range is a factor
six, compared with three or less elsewhere.  This deep minimum occurs over
several FUSE detector channels, and thus is unlikely to be a
detector, data processing, or tracking artifact.  Figure 2 shows our light
curves from the 1000\AA\ region and a longer wavelength region
(1090-1180\AA).  Similarly, G\"ansicke et al (1998) found a trend for the
amplitude of the modulation to be largest at their shortest wavelength
region which was centered on 1158\AA, but their HST observations did not
go into the shorter wavelength region covered by the FUSE data.

Mauche \& Raymond (1998) reported a flux range of a factor $\sim$3.3 at
1010\AA\ from their $ORFEUS~II$ data, and they did not note any
differences at other wavelengths.  They fit the continuum with a
two-temperature model, but there are deviations from it at wavelengths
shorter and longer than 1000\AA.  Hence, the effect of a larger 
variation
at $\sim$1000\AA\ may be present in their data as well, or there may be
some change in the FUV flux between their observations and ours.  They
have modeled the light curves with cyclotron components in the visible
region and with a localized hotspot on the white dwarf surface in the 
FUV.

Our wavelength-dependent light curve is not seen in the broad-line wings
when we take ratios of the two spectra.  This is demonstrated in Figure 3,
where the broad regions around the strong emission lines show the
shallower minima seen in the continuum below 950\AA\ and above 1050\AA. 
This suggests that there are very wide wings associated with the strong
emissions which fully overlap in the higher Lyman lines.  These features
are visible in the ratio of any two of the three FUSE orbits, all of which
have different binary phase coverage, but it is strongest in the ratio shown
in Figure 3, covering the largest continuum changes.  If the ratio in
the shortest wavelengths is affected by overlapping Lyman lines, it is
possible that the real continuum variation remains high at all wavelengths
below about 1000\AA.  Such a scenario is indicated schematically in Fig.\ 3
by the dotted line. This line is not calculated, but merely drawn in to show
a real continuum that would be feasible with broad-line blanketing, and 
compatible with the white dwarf hot-spot model.

\section{Emission Line Measurements and Changes}

The O VI lines are higher ionization lines than any emission observed 
in
the HST or optical ranges, and thus they are potentially of unique
interest.  They have been observed previously in two visits by $ORFEUS$
(in 1993 and 1996).  The 1996 data are superior and were taken when AM 
Her
was in a high state (see Mauche \& Raymond 1998).  The behavior of the 
O~VI
lines is similar to the He II line at 4686\AA\ in having sharp and
broad components with very different velocity amplitudes and phasing
(see e.g. Cowley and Crampton 1977). 

   Line velocities were measured by fitting a parabola to the entire feature,
and also by taking the flux centroid of the emission line. Values given are
the mean of the separate measures. The
broad component of the O VI doublet shows a velocity semi-amplitude
K$\sim$190 km s$^{-1}$ with the minimum velocity at $\Phi_{FUV}\sim0.9$ (or
approximately $\Phi_{mag}\sim0.0$), whereas the narrow component shows K
of only 62 km s$^{-1}$ with phasing about half a cycle later.  These
values can be directly compared to the measurements made by Mauche \&
Raymond who had fewer spectra but more evenly distributed around the
orbit.  Their amplitudes for the broad and narrow components of O VI were
K=412 and 57 km s$^{-1}$, respectively, with phasing very similar to what
we observe after taking into account the difference between FUV and
magnetic phases.

Additionally, there is a broad component of L$\beta$, easily seen on
either side of the airglow emission.  While blended with the airglow
emission, it appears to have a similar line
profile and to move with the broad O VI lines (see Fig. 4).  This broad
emission component is also present in the other strong Lyman lines where
it is too weak to measure.  Figure 2 includes our flux and velocity
measurements of L$\beta$.
The broad L$\beta$ line shows greater fractional change than the broad
O~VI, so that their ratio changes markedly at phases $\Phi_{FUV}$=0.4 to
0.5.

We also measured the flux and velocity of the C III emission line at
1175\AA.  This line is broad and has no narrow component significantly
different from the noise level.  We have also summed the spectra in groups
of three to increase the signal-to-noise in the line profile.  Its overall
profile changes strongly, as shown in Figure 5.  Using the C III
velocities from all 19 spectra (to retain the maximum phase resolution), 
we obtained the sine-curve fit given in
Table 2.  Both the semi-amplitude (K=311 km s$^{-1}$) and the mean value
(+50 km s$^{-1}$) are significantly different from the fit found for the
broad O VI lines.  G\"ansicke et al.\ found C III 1176\AA\ to have a
semi-amplitude of 540 km s$^{-1}$ with the minimum velocity occurring
about 0.1P later than we observe.  However, their HST data had more
complete phase coverage, so these two C III lines probably behave
similarly.

The other emission lines were weaker and contaminated with airglow
emissions, so we were not able to measure them.  However, Fig.\ 3 suggests
the presence of other possible emission features if the dips are all caused
by line broad line emission that has a lower flux variation than the continuum.
This idea is reinforced by the fact that these wavelengths correspond to those
of possible emission lines.  Along the bottom of this figure we have marked
the positions of the identified  lines we measured, as well as the stronger 
S~IV line positions - all of which lie at dips in the ratio plot. Thus, we
conclude that several weak S IV emission features may be present that do
not show up clearly in either the individual or summed spectra.

While the phasing of FUV continuum are in good agreement with those of
Mauche \& Raymond and with the UV data of G\"ansicke et al., the velocity
amplitudes of the broad lines differs, as described above.  Our amplitudes
are less well-defined because of our more limited phase coverage, but they
do appear to cover the velocity extremes fairly well.

\begin{deluxetable}{llcccr}
\tablecaption{Sine Curve Fits to Measurements}
\tablehead{\colhead{Quantity} &\colhead{Minimum}
&\colhead{$\Delta\phi$} &\colhead{Semi-amplitude} &\colhead{Mean}
&\colhead{Mean} \\
&\colhead{HJD 2451700+} &\colhead{phase\tablenotemark{a}} &\colhead{K}
&\colhead{Value} &\colhead{Error} }
\startdata
~~~~\underbar{Flux} & & & \underbar{mag} & 
\underbar{erg/cm$^{2}$/sec/\AA} 
& \underbar{mag} \nl
980--1090\AA\  &8.4594$\pm$0.0011 &(0.0) & 0.65$\pm$0.05
&2.17$\times10^{-12}$ &0.05 \nl
994--1020\AA\  &8.4585$\pm$0.0011 &0.993$\pm$0.009 &1.20$\pm$0.05
&1.84$\times10^{-13}$  &0.08  \nl
1090--1180\AA\ &8.4619$\pm$0.0011 &0.019$\pm$0.009 &0.61$\pm$0.05
&2.17$\times10^{-13}$ &0.05 \nl
\nl
~~\underbar{Velocity} & &  & \underbar{km s$^{-1}$}
& \underbar{km s$^{-1}$} & \underbar{km s$^{-1}$}   \nl
O VI broad &8.4479$\pm$0.0041 &0.910$\pm$0.03 &190$\pm$15
&$-$30$\pm$14 &79 \nl
O VI peak &8.3828$\pm$0.0014 &0.406$\pm$0.01 &62$\pm$2
&16$\pm$2 &9 \nl
C III broad &8.4416$\pm$0.005 &0.862$\pm$0.04 &311$\pm$38
&50$\pm$31 &125 \nl
\enddata
\tablenotetext{a}{Phase of minimum flux or velocity,
with respect to FUV ephemeris }
\end{deluxetable}

\section{Absorption Lines}

As noted earlier, H$_2$ absorption is weak, indicating there is
little H$_2$ along the line of sight.  In the supersoft binary system QR
And where H$_2$ was particularly strong, we presented evidence that this
absorption was at least partially due to circumbinary material.  By
comparison, this suggests that little cool gas is leaving the AM Her
system.  However, we do see a clear set of narrow H absorption lines (see
Figure 1).  Within the precision of the FUSE wavelength scale at the
shortest wavelengths, these H absorptions all show the same velocity and
do not change with time.  Their mean velocity is $-36\pm$8 km s$^{-1}$,
which can be compared to the systemic velocity of $-19$ km s$^{-1}$
(Young, Schneider, \& Shectman 1981). We note however, that our Calfuse
1.9.9 extractions gave different results in the region than the version
1.6.9 pipeline data. 
We regard ours as better as a) the version 1.6.9 radial velocities range
systematically with wavelength while the later version 1.9.9 values do not.
The FUSE wavelength scale precision is not well established at the shortward
limit, but it is possible that there is a systematic velocity error of
order $\sim$20 km s$^{-1}$.

The Lyman absorption lines between 918 and 950\AA\ were measured for
equivalent width.  Noting again a difference between our extractions and
the pipeline, we find the absorptions have a mean EW of 0.26$\pm$0.04\AA.
There is no significant change with binary phase.  Thus, the absorption
appears to arise outside the system, and may be normal interstellar gas.

There are several weak to moderate absorption features seen in all spectra
in the wavelength range 1036 to 1041\AA, where we have good signal and
are interested in contamination of the O VI emission.  
We identify the strongest as the
C II doublet at 1036.3, 1037.0\AA. Other weak absorptions match the 
H$_2$ spectrum at the same velocity (close to zero, subject to the FUSE
wavelength scale uncertainty), and also O I.
We do not see any changes in these absorptions with binary phase.  
They are weak enough that they do not affect the O VI line measures
significantly.

\section{Summary}

A general model of the system has been discussed and developed by a 
number
of authors (e.g. G\"ansicke et al. 1998 and Mauche \& Raymond 1998 are
recent examples).  There is mass transfer along magnetic field lines
within the system, principally on to one pole of the white dwarf, at 
least in
the system high state.  This
pole is hot and causes the X-ray and UV flux variations as the system
rotates.  The optical-NIR light changes are caused by a combination of 
the
temperature gradients on both the donor star and the white dwarf.  The 
broad
emission lines arise along the path of the gas stream, while the narrow
emission lines appear to be located near the cusp of the donor star's 
Roche
lobe.  The FUSE data are consistent with this picture, while providing
some new details and aspects of the complex system.  Our principal 
results
are as follows.

1. The broad O VI emission lines have the same phasing as the other 
broad
emissions, but appear to have a different velocity amplitude from that
seen in 1996 in the ORFEUS data.  Changes in other broad-line 
amplitudes
have been noted (e.g. by Crosa et al. 1981 for He II 4686\AA), and
presumably indicate variations in the gas stream ionisation and
kinematics, as the mass transfer rate varies.

2. The narrow O VI emission lines also behave much like the other narrow
emissions.  Our data agree with the ORFEUS results in velocity, but the
higher FUSE resolution gives a very different separation of flux between
the broad and narrow components.  While we do not cover the entire orbit,
we find the narrow line to have less that 1/4 the broad line flux
at all times and that the narrow line is very weak at FUV light minimum.

The velocities of the narrow lines seem more stable over the history of
various investigations and support the standard interpretation that they
arise at the Roche lobe cusp.  The range of ionization seen (from Ca II to
O VI) is very large, and it is clear that some mechanisms other than
photoionization are involved.  If we examine the velocity amplitudes for
all the lines, there is a trend with ionization potential running from
values near 100 km s$^{-1}$ for Ca II and H I to about 60 km s$^{-1}$
for O VI and N V.  He I, C
II, He II, Si IV, N III are part of the intermediate sequence.  As
proposed by G\"ansicke et al., this suggests an ionization gradient moving
away from the Roche lobe cusp towards the system center of mass.

3. The FUV light curve shows strong changes over the FUSE wavelength range.
This is compatible with the model of a hot spot on the white dwarf if
a significant part of the FUV flux arises as broad line emission from all
the lines detected and where the higher Lyman lines overlap each other.

4. We find sharp H I Lyman absorptions with a constant radial velocity
of $-$36 km s$^{-1}$. 
Weak H$_2$ absorption appears to be present, which does not change with
binary phase, and has the same velocity as other interstellar absorptions.  
AM Her's H$_2$ absorption is much
weaker than that in the X-ray binary QR And, which is at a similar
distance from us.

\acknowledgments

The authors acknowledge use of the AAVSO web site and thank the many
variable star observers for their contributions to this dataset.  APC 
also
acknowledges her support from NASA for this work.

\clearpage

\begin{figure}
\caption{Two AM Her spectra summed from full FUSE orbits (2 and 3),
covering FUV phases near maximum and minimum light.  The dashed lines
show the zero values for the upper spectra. The strong emission
lines are marked and identified.  The strong narrow emissions (with the
exception of O VI) are airglow.  Note the broad emission profile 
changes
and the narrow Lyman absorptions. }
\end{figure}

\begin{figure}
\caption{Measurements from 19 equal-exposure FUSE spectra plotted on 
FUV
phases (see text for conversion to binary or magnetic phases).
In the valocity panels the open circles are the broad O VI values, the
dots the O VI peaks, the asterisks the C III, and the encircled crosses 
are
L$\beta$.  Phase zero
is the minimum FUV light from a sine fit to the LiF1a channel data.  
Note
the deeper minimum at $\sim$1000\AA\ than at longer and shorter
wavelengths.  Other panels show radial velocities from emission-line
centroids and total flux in emission-line features.}
\end{figure}

\begin{figure}
\caption{Ratio of FUSE orbit spectra taken near maximum and minimum FUV
light (as shown in Fig.\ 1), after airglow removal and heavy smoothing.
Note the strong increase in continuum ratio below 1050\AA\ and the 
lower
amplitude seen in broad `wings' around the emission lines.  The 
tickmarks
show the positions of emission lines, including weak S IV features.  It 
is
possible the ratio is low below 960\AA\ because of overlapping Lyman
series `wings'.  If so, then the real continuum variation may remain 
high
at all wavelengths below about 1000\AA.  The dotted line indicates such 
a
possible interpretation, as discussed in the text.  }
\end{figure}

\begin{figure}
\caption{The full sequence of O VI and L$\beta$ emissions (with airglow
removed).  The spectra are smoothed to reduce noise but that has 
broadened
the narrow emissions.
Plots are offset by 5$\times10^{-13}$ erg cm$^{-2}$ sec$^{-1}$
\AA$^{-1}$ in flux.  See Table 1 for the FUV phases for each spectrum. 
}
\end{figure}

\begin{figure}
\caption{Variations in the C III 1175\AA\ line.  Three or four spectra
have been combined in each plot to show the changes more clearly.
Continua, interpolated from adjoining parts of the spectrum, are dotted
in. The vertical bars show the velocity centroids of the emission 
lines.
Plots are offset in flux by $2\times10^{-13}$ erg cm$^{-2}$
sec$^{-1}$, and their mean FUV phases are given.  Figure 2 plots 
measurements
from these profiles. }
\end{figure}

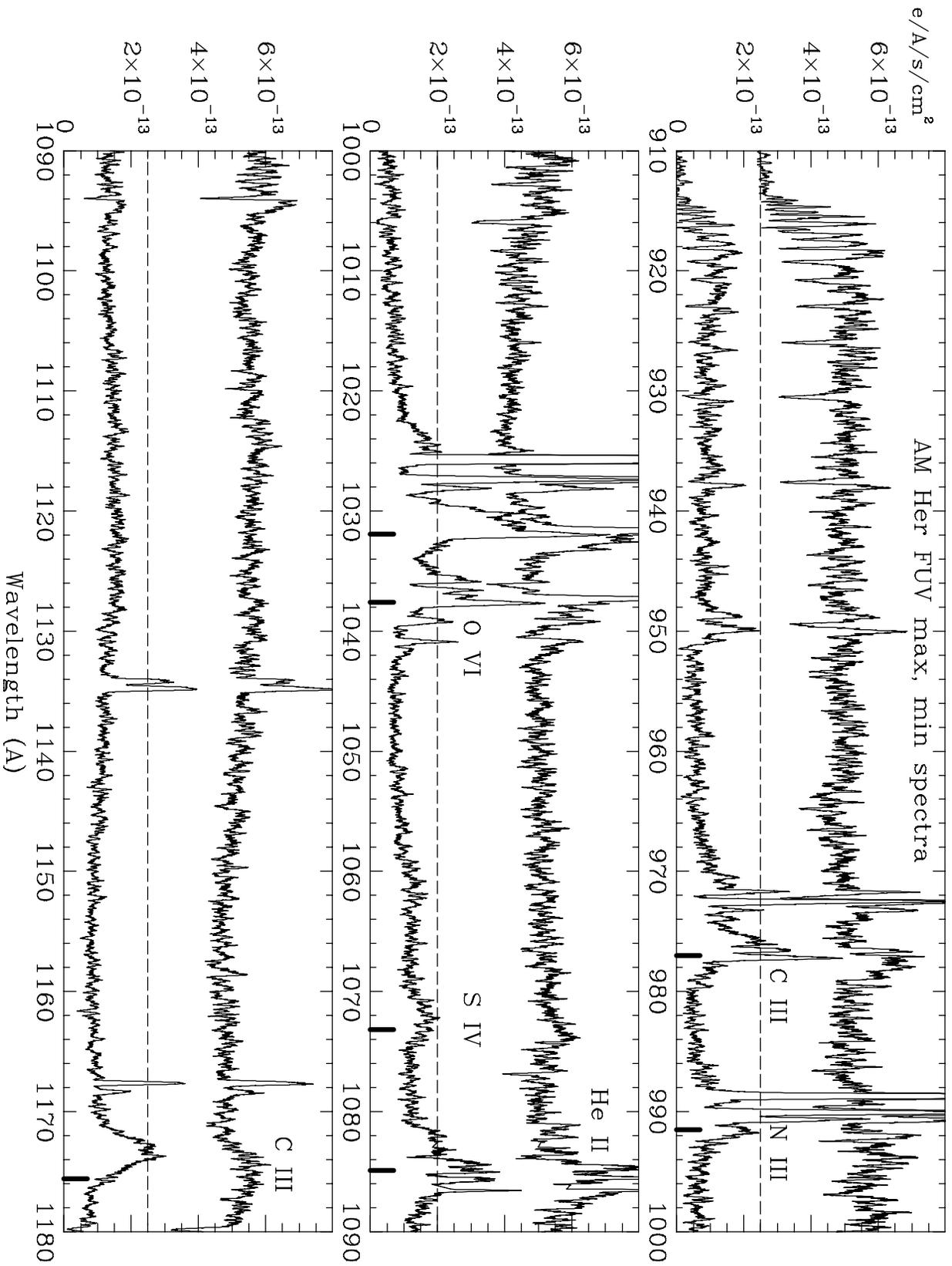
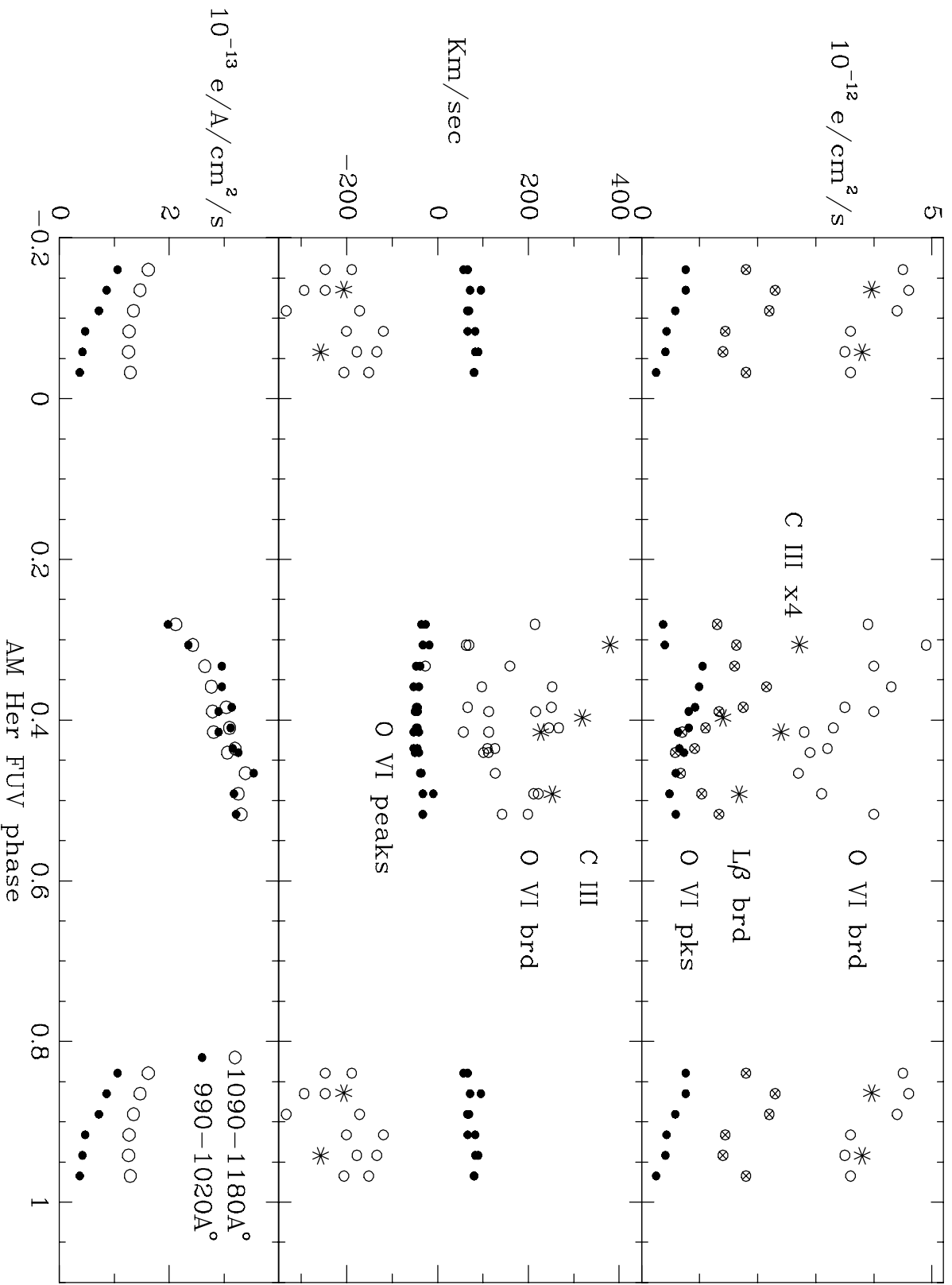
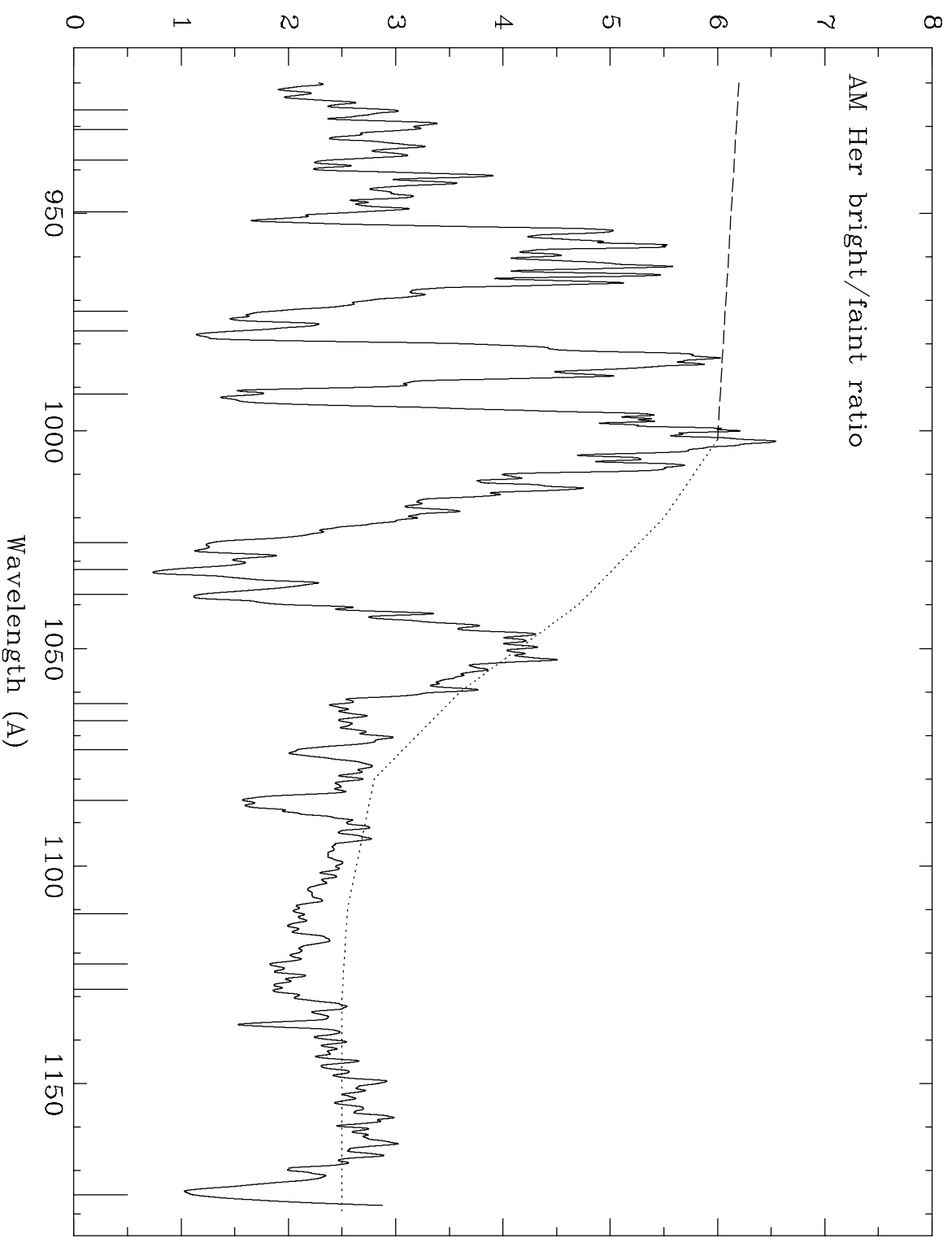
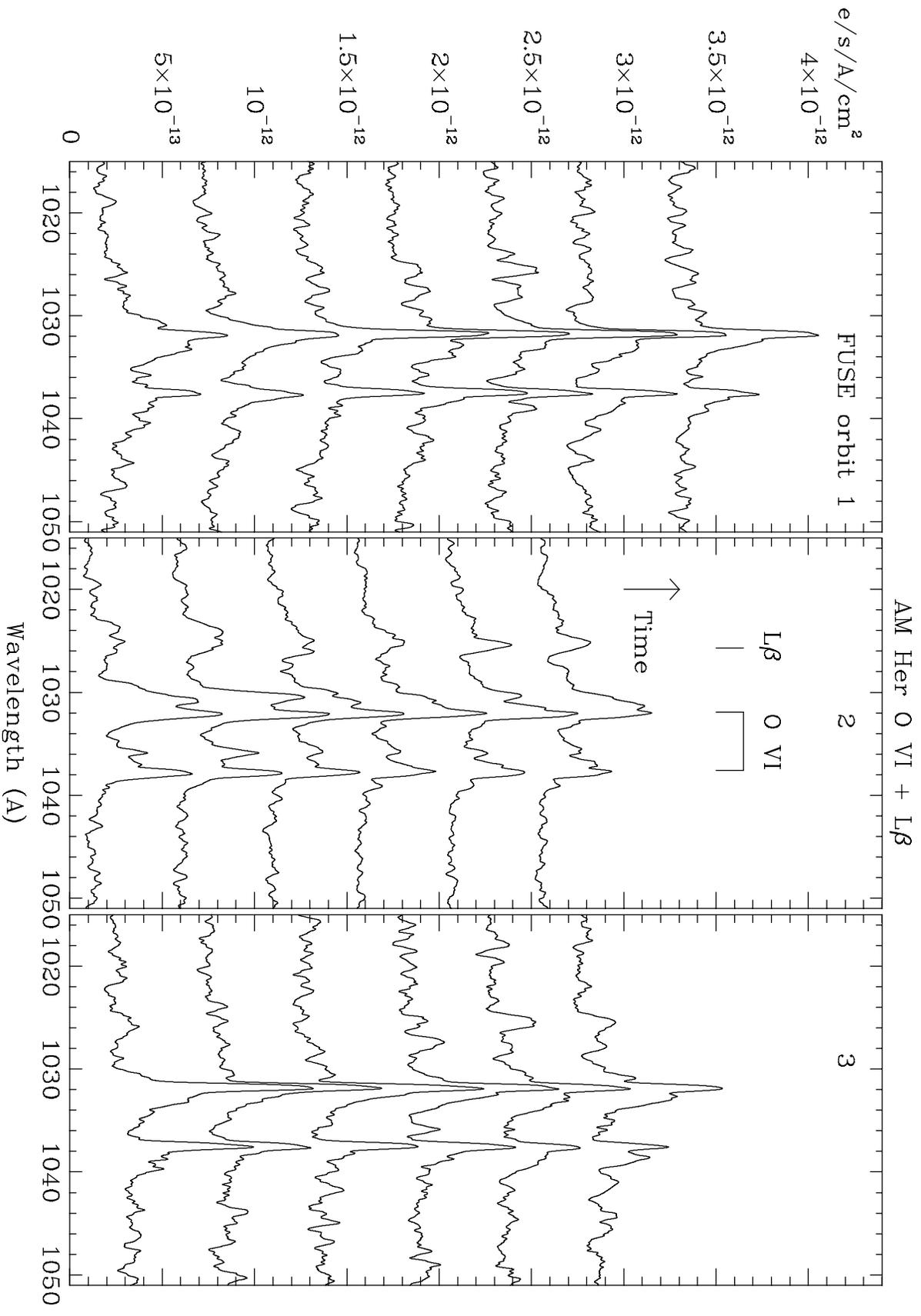
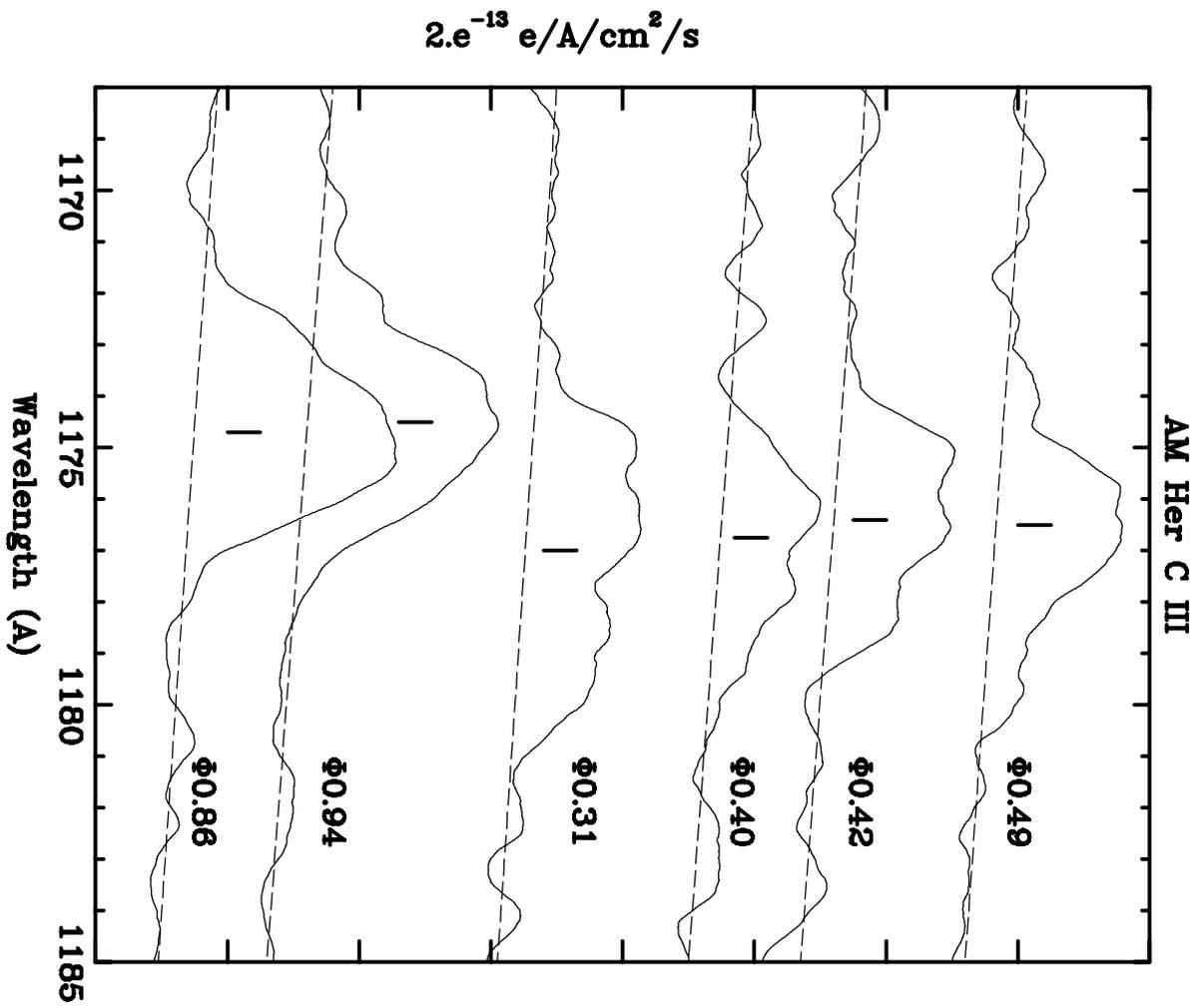
\end{document}